
\font\titlefont = cmr10 scaled \magstep2
\magnification=\magstep1
\vsize=20truecm
\voffset=1.75truecm
\hsize=14truecm
\hoffset=1.75truecm
\baselineskip=20pt

\settabs 18 \columns

\def\b{\bigskip}
\def\bb{\bigskip\bigskip}

\def\ce{\centerline}

\def\no{\noindent}




 \rightline{ UMDHEP 94-04}
\rightline{ July 1993}
\bb

\b
\ce{\titlefont{Cosmological Constraint on the Scale of the}}
\ce{\titlefont{ Supersymmetric Singlet Majoron
    \footnote\dag{\rm{ Work supported by a grant from the National
          Science Foundation}} }}
\bb
\ce{\bf{R.N. Mohapatra and X. Zhang
   }}

\ce{\it{ Department of Physics and Astronomy}}
\ce{\it{University of Maryland}}
\ce{\it{ College Park, MD 20742 }}
\b
\ce{\bf Abstract}

\no
In a supersymmetric theory with a spontaneously broken global symmetry,
$G$, if the scale of supersymmetry breaking, $M_s$ is smaller than
the scale $M_G$ of the global symmetry, the Nambu-Goldstone boson,
$\chi$ is accompanied by two massive superpartners
( a fermionic, $\Psi_\chi$
and a scalar boson, $\sigma_\chi$ )
with mass of order $M_s$. Cosmological considerations imply
stringent constraints on the couplings of $\Psi_\chi$
and
$\sigma_\chi$. Application of these considerations to the
supersymmetric singlet Majoron (SUSYSM) model
leads to an upper limit on the
scale $V_{BL}$ of global
$U(1)_{B-L}$ symmetry to be
$\leq 10^{4}$GeV,
for reasonable values
of parameters in the theory.

 \filbreak

It is widely believed that the fundamental particle interactions may be
supersymmetric beyond the TeV scale in order to solve the problem of mass
hierachy between the Fermi and Planck scales. There are also various
reasons for considering the existence of global symmetries
(perhaps only U(1) symmtries) of nature which are spontaneously broken.
Examples are that of $U(1)_{PQ}$ symmetry needed to solve
the strong CP problem[1] or the
$U(1)_{B-L}$ symmetry, widely discussed in understanding
the nature of massive neutrinos[2]. It is a conceivable  that a complete
theory of nature
is one that encompasses a high energy supersymmetry as well as
a spontaneously broken global symmetry. It is, then, of interest to see
if cosmological considerations imply any new constraints on such
theories.

In a supersymmetric theory with a spontaneously broken global symmetry, G,
if the scale, $M_G$ of the global symmetry breaking is much larger than the
 scale $M_s$ of supersymmetry breaking, then the effective theory for
$\mu \ll M_G$ contains a (or a set of) massless Nambu-Goldstone boson(s)
corresponding to the broken generator(s) of $G$ and
its (their) superpartners, which have masses of order, $M_s$.
Specializing to the case where $G$ is a $U(1)$ symmetry, the Nambu-Goldstone
boson ($\chi$) will be accompanied by a 2-component neutral fermion
($\Psi_\chi$) and a scalar
boson
$\sigma_\chi$.
Cosmological constraints on
these particles for the case of
 SUSY $U(1)_{PQ}$ models have been widely discussed
in the literature[3]. In this note, we focus our attention on a supersymmetric
theory where global $U(1)_{B-L}$ symmetry is spontaneously
broken[4] by a $SU(2)_L \times U(1)_Y$ singlet field. In such theories, the
scale, $V_{BL}$ of the B-L symmetry breaking is connected to the neutrino
mass via the see-saw mechanism. Any information on $V_{BL}$
will therefore provide information on the nature of light
neutrino masses. So far, only very weak constraints
(${\it i.e.}~ V_{BL} \geq O(100~{\rm GeV})$) can be deduced from the
astrophysics of red giant stars[4]. Recently, some more
constraints on $V_{BL}$
have been deduced if
one assumes that the Planck scale effects break the global B-L symmetry
by dimension 5 operators[5-8]. In this letter, we discuss constraints
on $V_{BL}$ that
arise if the singlet Majoron model is made supersymmetric. There are two
possible points of view: one is to consider a minimal extension of the SUSY
standard model (MSSM) by including the right-handed neutrino superfield
$\nu^{c}$ and a singlet field $S$ with
$L = +2$[9,10]. In this model, scale
$V_{BL}$ and
$M_s$ are necessarily of the same order in order for
spontaneous breaking of $U(1)_{B-L}$ to occur. There is then spontaneous
breaking of R-parity[11] in this model. The second possibility, not discussed
in the literature to date, is to consider $V_{BL} \gg M_s$, which, as we
discuss below, necessarily requires
two more
$SU(2)_L \times U(1)_Y$ singlet superfield in addition to
$\nu^c$ and
$S$. We will show that in this class of theories,
cosmological constraints imply that
$V_{BL} \leq 10^{4}$GeV.

\no {\bf { The SUSYSM Model:}}
 In order to derive the above mentioned constraint on $V_{BL}$,
we will work with a generic supersymmetric theory
where $V_{BL} \gg M_s$. The superfield
content of the model along with their transformations
under $SU(2)_L \times U(1)_Y$ as well as B-L global symmetry
is shown in table I.

 The superpotential
 for the model can be written as a sum of
two terms:
$${
W = W_0 + W_1~~, }$$

\vbox{\tabskip=0pt \offinterlineskip
\def\tablerule{\noalign{\hrule}}
\halign to300pt {\strut#& \vrule#\tabskip=1em plus2em&
  \hfil#& \vrule#& \hfil#\hfil& \vrule#&
    \hfil#& \vrule# \tabskip=0pt \cr \tablerule

&& Superfield && $SU(2)_L \times U(1)_Y$  && $U(1)_{B-L}$
                                                               &\cr \tablerule
&& Q   &&  ( 2, 1/3 ) &&  +1/3   &\cr \tablerule
&& $u^c $ && (1, -4/3 ) && -1/3 &\cr \tablerule
&& $d^c$  && (1,  +2/3) && -1/3 &\cr \tablerule
&& L &&   (2, -1) &&   -1 &\cr \tablerule
&& $e^c $ && (1, +2) && +1 &\cr \tablerule
&& $\nu^c $ && (1, 0) && -1 &\cr \tablerule
&& $H_u$ && (2, +1) && 0 &\cr \tablerule
&& $H_d$ && (2, -1) && 0 &\cr \tablerule
&& S && (1, 0) && +2 &\cr \tablerule
&& $S^\prime$ && (1, 0) && -2  &\cr \tablerule
&& Z && (1, 0) && 0 &\cr \tablerule \noalign{\smallskip}
& \multispan{7} Table I. Superfields and their transformations
\hfil \cr
&\multispan{7} under
$SU(2)_L \times U(1)_Y \times U(1)_{B-L}$ \hfil \cr }}

\bb
\b

\no where
$${
\eqalign{
W_0 = & h_u Q H_u u^c + h_d Q H_d d^c + h_e L H_d e^c + h_\nu L H_u \nu^c \cr
      & + f \nu^c \nu^c S + \mu H_u H_d ~~, \cr }
} $$
\no and
$${W_1 = \lambda ( S S^\prime - M_1^2 ) Z~~. } \eqno(1)$$
\no We choose $M_1 \gg V_{WK}$, the electroweak scale.
The vanishing of F-terms at the scale $M_1$ (which is of order $V_{BL}$)
in order to maintain supersymmetry is satisfied by
$${
<S> = V_S ; ~~ <S^\prime > = V_{S^\prime}; ~~ <Z> = 0 ; ~~
<\tilde{\nu^c}> = 0 , ~~
 }\eqno(2)$$
\no with $V_S V_{S^\prime} = M_1^2$.
 $V_S$ and
$V_{S^\prime}$
will be
assumed to be of the same order ($ {\it i.e.}~~V_{BL} \simeq M_1
\gg V_{WK}$). It is easy to work out the particle spectrum for
$\mu \ll V_{BL}$. Apart from the quark, lepton
and Higgs fields $H_{u, d}$, there are three massless fields:

\no {\bf i)} the Majoron $\chi$:
$${
\chi = {V_S \chi_S - V_{S^\prime} \chi_{S^\prime} \over {
\sqrt{ V_S^2 + V_{S^\prime}^2 } } } ~~;      }\eqno(3)$$

\no {\bf ii)} Majorino $\Psi_\chi$:
$${
\Psi_\chi =
    { V_s \Psi_S - V_{S^\prime} \Psi_{S^\prime} \over
   { \sqrt{ V_S^2 + V_{S^\prime}^2 } } } ~~;} \eqno(4)$$

\no{\bf iii)} Smajoron $\sigma_\chi$:

$$
{ \sigma_\chi = {
V_S \sigma_S - V_{S^\prime} \sigma_{S^\prime} \over {
\sqrt{V_S^2 + V_{S^\prime}^2 } } }~~, }\eqno(5)$$
\no where we have written the superfield as:
$${
S = {1\over \sqrt2}( \sigma_S + i \chi_S )
  + {\sqrt2} \theta \Psi_S + \theta^2 F_S ~~, }\eqno(6)$$
\no and similarly for $S^\prime$. In the absence of explicit
supersymmetry breaking terms, all of the light fields
($Q,~L,~H_{u, d},~ \chi ,~ \sigma_\chi ,~ \Psi_\chi$) are massless
and the electroweak symmetry is unbroken. In order
to make the theory realistic, we will add to the lagrangian
the soft SUSY breaking (but B-L conserving) terms which have the form:
$${
\eqalign{
   V_s =& \sum_{a} \mu^2_a \phi^\dagger_a \phi_a + \sum_{a}
A_a \int d^2\theta \theta^2 W_a \cr
       &+ \sum_{a} m_{\lambda_a} \lambda_a \lambda_a + h.c ~~,\cr}
}\eqno(7)$$

\no $\phi_a$ goes over all scalar superpartner of light fields and
$\lambda_a$ are gaugino fields.
$W_a$ denotes each term in the superpotential. The origin of
the soft SUSY breaking potential, $V_s$
is irrelevant to our subsequent discussion. We will assume
the parameters of $V_s$ to be such that
they induce
electroweak symmetry breaking
${\it i.e.} <H_u> = V_u ~~{\rm and}~~ <H_d>= V_d$ as usual.
Two immediate consequences follow from eq.7. Firstly,
that since the terms in $V_s$
 respect B-L symmetry, the Majoron, $\chi$
remains massless. On the other hand, the SUSY breaking
terms impart a mass to $\sigma_\chi$ of order $M_s$. We will
assume $M_s \simeq 1$TeV in the subsequent discussion.

In order to discuss the Majorino mass, we first note that once
supersymmetry is broken, the scalar component of the singlet field Z
acquires a vev: $<z> \simeq M_s ~ V_S ~ V_{S^\prime}~/ M^2_z ~
\sim M_s$. This then gives a tree level mass to the Majorino of order
$M_s ~ {\it i.e} ~ m_{\Psi_\chi} \simeq M_s$, which can be of the order
of a TeV.
\b

\no {\bf { Cosmological Constraints:}}
   In generic SUSY models of the type we are considering, the dominant
interactions of ($\chi,~~ \Psi_{\chi},~~\sigma_{\chi}$) are with the superheavy
particles, such as $\nu^c , ~~ (V_{S^\prime}S + V_{S} S^\prime )$ superfields.
Any interaction with light particles arises via the coupling
$h_\nu L \nu^c H_d$ and is therefore suppressed by the inverse powers
of $V_{BL}$. This observation has important implications for cosmology, since
for the epochs of the universe below the temperature $T< V_{BL}$, all the
heavy particles annihilate and disappear. As a result, the interactions
(scatterings as well as decays) of the particles $\Psi_\chi$
and
$\sigma_\chi$ become very weak. In order to study their impact on the
evolution of the universe, we have to find the temperature, at which
$\Psi_\chi$
and
$\sigma_\chi$ decouple from the rest of the particles since
this determines their abundance at subsequent epoch until nucleosynthesis
temperature. If this aboundance is significant, we have to
find their life time to study their impact on nucleosynthesis.

\no {\it {a) Determination of decoupling temperature:}}

We are interested in the case where $V_{BL} \gg V_{WK}$.
For $T \geq V_{BL}$, all particles are massless and are in equilibrium.
 For
$T < V_{BL}$, the dominant effective interactions that can keep the
$( \chi , ~~ \sigma_\chi , ~~\Psi_\chi )$ in equilibrium with leptons and
quarks
is of the form:

$${\eqalign{
    {\cal L}_{eff} \simeq & {\epsilon_1 \over V_{BL}^2}
  {( \partial_\mu \chi )}^2 {\tilde l_a} {\tilde l_b} +
  { \epsilon_2 \over V_{BL}^2 } {( \partial_\mu \sigma_\chi )}^2
     {\tilde l_a} {\tilde l_b} \cr
                          & + {\epsilon_3 \over V_{BL}^2} {\overline \Psi_\chi}
                 \gamma_\mu \partial^\mu \Psi_\chi {\tilde l_a}{\tilde l_b}
                      + ...... \cr}
}\eqno(8)$$
\no The .... stands for Higgs fields replacing $\tilde l$.
 These interactions arise from the D-type terms
 induced at the one loop, therefore
$\epsilon_i$ are expected to be small, typically\footnote{[F.1]}{
Here as well as in the rest of the paper, we will
only carry out the order of magnitude estimates for parameters. More precise
statements than this
would require detailed structure of the model, which is not important
at the present stage.}
$${
\epsilon_i \simeq { h_\nu^2 \over {16\pi^2} } ~~ ,}\eqno(9)
$$
\no where $h_\nu$ is the coupling of $\nu^c$ to the heaviest light neutrino.
The order of magnitude of the decoupling temperature
for $\Psi_\chi$
and $\sigma_\chi$
is
then determined by the condition:
$${
{\epsilon_i^2 ~ T^5 \over V_{BL}^4 } \leq {[g_{*}(T)]}^{1/2} {T^2 \over M_{pl}}
{}~~~. }\eqno(10)$$
\no This leads to
$${
T_D \leq 10^{4.3}~ V_{BL}~ {( {V_{BL}\over M_{pl} })}^{1/3}~
                {( {10^{-6}~ \over \epsilon_i } )}^{2/3}
                ~ {\rm GeV}. }\eqno(11)$$
\no In the specific model under consideration, since
the neutrinos cannot decay fast enough[12], we would expect their masses
to satisfy the cosmological constraint, $m_\nu \leq 40$ eV[13], which
would imply that,
$${
h_\nu^2 \leq {40~({\rm eV})~ f ~ V_{BL} \over V_{WK}^2 } ~~. }\eqno(12)$$
Using this equation and eq.11 we find (all parameters in GeV units)
$${
T_D \leq  10^{9.6}~ {( { V_{BL}^2 \over {M_{pl} ~f^2 } })}^{1/3}~~
                     {( {40{\rm eV}
                      \over m_\nu } )}^{2/3}
              ~~{\rm GeV} ~~; }\eqno(13.a)$$
\no or

$${ T_D \sim V_{BL} ~~, }\eqno(13.b)$$

\no whichever is lower.

For $T< T_D$ until the particles
$\sigma_\chi ~{\rm and}~ \Psi_\chi$
 decay, their number density
 decreases only due to
expansion of the universe $( n \sim T^3 )$
{\bf except} at various annihilation thresholds for
massive particles. Therefore, for temperature $T( \tau )< T < T_D$
(where $T( \tau )$ is the temperature at the decay epoch of these particles),
$$
{
{n( \sigma_\chi ) \over n_{\gamma} } {\bigm |}_T \simeq
         {g_{*} (T) \over g_{*} (T_D) }  ~~~.}
\eqno(14)$$
\no Since $M_{\sigma_\chi} \simeq M_s \simeq 1$TeV, the
Smajoron will dominate the mass density of the universe below
approximately $T \simeq 10$GeV
 and will completely upset the discussions of nucleosynthesis,
if it is stable.
In order to maintain our present excellent understanding of the
 nucleosynthesis[14],
we demand
that the heavy Majorino and Smajoron decay before $t \leq 10^{-2}~
{\it sec.}$

\no {\it {b) Decay of Majorino ($\Psi_\chi$) and
Smajoron ($\sigma_\chi$) }}:
Let us first consider the Smajoron decay. Above the electroweak phase
transition
temperature, the $\sigma_\chi$ is absolutely stable. For $T < V_{WK}$,
however,
 the decay $\sigma_\chi \rightarrow \nu ~\nu$ can occur with
a coupling strength of order due to non-zero
Dirac mass of the neutrino:
$${
g_{( \sigma_\chi \rightarrow \nu \nu )} \simeq { h_\nu^2\over f} {(
{V_{WK}\over
    V_{BL} })}^2
{}~~. }\eqno(15)$$

\no In eq.15, we have kept only the heaviest of the light
left-handed neutrinos.
 Requiring $\tau_{\sigma_\chi} < 10^{-2} {\it sec.}$[15], we then find,
$${
V_{BL} \leq 10^{8}~ h_\nu~ {({ m_{\sigma_\chi}~ \over{ 1{\rm TeV}~f^2 }}
 )}^{1/4}
                 ~~{\rm GeV}~~~~. }\eqno(16)$$

\no Again
 using the see-saw formula for neutrino masses
$m_\nu \simeq h_\nu^2 V_{WK}^2 / f V_{BL} $, we have that
$${
V_{BL} \leq 10^4 {( {m_\nu \over {40 {\rm eV}} })}
                   ~{( {m_{\sigma_\chi} \over {1 {\rm TeV}} })}^{1/2}
                      ~~~{\rm GeV}
   ~~~.}\eqno(17)$$

\no As pointed out above that
 in the minimal singlet Majoron model the neutrinos are likely
to have a long life time[12],
  we
use the cosmological upper limit of
40 eV on the mass
 $m_\nu$ of stable neutrino[13] and get
$V_{BL} \leq 10^4$ GeV. In non-minimal Majoron models, neutrinos may
be unstable and therefore may be heavier than 40 eV. The upper limit on
$V_{BL}$ is then less stringent.

Turning now to the decay of the Majorino, let us assume that,
$m_{\Psi_\chi} \geq m_{\tilde H}$, where
$\tilde H$ is the lightest neutralino. Present data therefore implies
that $m_{\Psi_\chi}$ should be
in the 100 GeV range (or higher). The case
$m_{\Psi_\chi} \leq m_{\tilde H}$ is discussed later on. The dominant decay
of $\Psi_\chi$
is non-vanishing only for $T < V_{WK}$,
and is a tree
-level process mediated by virtual $\tilde{\nu_c}$ exchange, leading to
$\nu ~ \nu ~{\tilde H}$ as a final state.
 The strength of the
$\Psi_\chi \rightarrow \nu~\nu~{\tilde H}$ coupling is
$${
g_{\Psi_\chi \rightarrow \nu \nu {\tilde H} }
\simeq { h_\nu^2~ f~ V_{WK} \over {2 {\sqrt2}V_{BL}^3 } }
{}~~~~.}\eqno(18)$$

\no Again requiring $\tau_{\Psi_\chi} \leq 10^{-2}~{\it sec.}$, we get,
using the see-saw formula
$${
V_{BL} \leq 4\times 10^{3}
    {( {m_\nu \over {40 {\rm eV}} })}^{1/2} ~f~ {( {m_{\Psi_\chi} \over
            {1 {\rm TeV}} })}^{5/4} ~~{\rm GeV}~~~.}\eqno(19)$$
\no This bound is of the same order as in eq.17.

The case where $m_{\Psi_\chi} < m_{\tilde H}$ is interesting because, in
 this case, either the abundance
$\Psi_\chi$ must be reduced by annihilation process or
the SUSYSM model is ruled out. Note that, since in this case, R-parity
is an exact symmetry, the decay of $\Psi_\chi$
is forbidden,
if it is
lighter than the lightest neutralino.
And its annihilation channels are also inefficient if $V_{BL} > $ TeV (see
eqs.13). Therefore, if $m_{\Psi_\chi} < m_{\tilde H}$, our conclusion is that
$V_{BL}$ must be less than a TeV.

A general concern in the case of late decaying of heavy particles is the
possible dilution of baryon to entropy ratio below the observed value. This
question has been analized in detail by Scherrer
and Turner in ref.15.
For $m_{\sigma_\chi , \Psi_\chi} \leq 1$ TeV,
and lifetime
$\tau \leq 10^{-2} {\it sec.}$ considered here,
they have shown that neither nucleosynthesis nor
baryon to photon ratio is effected by their late decay.

In summary, we have found that
if spontaneous breaking of global B-L symmetry
occurs in a supersymmetric model,
the scale $V_{BL}$ is likely to be in the TeV range. It is therefore
quite likely to manifest itself in rare decay processes.
It is also worth emphasizing that, while we have carried out
our discussion using the minimal singlet Majoron
model, all our considerations hold for more elaborate versions of it.

\bb
\b

We thank K.S. Babu for a useful remark.

\b
\bb

\ce{\bf References}
\b
\item{[1]}
R.D. Peccei, in {\bf {CP Violation}}, ed. by C. Jarlskog,
( World Scientific,
Singapore
1989 ); J.E. Kim, Phys. Rept. {\bf 150}, 1 (1987);
H.Y. Cheng, Phys. Rept. {\bf 158}, 1 (1988).

\item{[2]}R.N. Mohapatra and
  P.B. Pal, {\bf {Massive
Neutrinos in Physics and Astrophysics}} (World Scientific, 1990).

\item{[3]}
K. Tamvakis and D. Wyler, Phys. Lett. {\bf B112}, 451 (1982);
J.E. Kim, Phys. Lett. {\bf B136}, 378 (1984);
J. Nieves, Phys. Rev. {\bf D33}, 1762 (1986);
P. Moxhay and K. Yamamoto, Phys. Lett. {\bf B151}, 363 (1985);
K. Rajagopal, M.S. Turner and F. Wilczek, Nucl. Phys.
{\bf B358}, 447 (1991);
D.H. Lyth, Lancaster Preprint (1993).

\item{[4]} Y. Chikashige, R.N. Mohapatra and R.D. Peccei,
Phys. Lett. $\bf 98B$, 265 (1981).

\item{[5]}
E. Akhmedov, Z. Berezhiani, R.N. Mohapatra
and G. Senjanovi\'c, Phys. Lett. {\bf B299}, 90 (1993).

\item{[6]}
I. Rothstein, K.S. Babu and D. Seckel,~~~~~~~~ Bartol Research Institute
                                  Preprint (to be published).

\item{[7]}J. Cline, K. Kainulainen and K. Olive, UMN-TH-1113-93,
TPI-MINN-93/13-T,
     (to be published).

\item{[8]}R.N. Mohapatra and X. Zhang, Phys. Rev. {\bf D48 } in press
         (1993).

\item{[9]}A. Masiero and J.W.F. Valle, Phys. Lett. {\bf B251}, 273 (1990).

\item{[10]}G.F. Giudice, A. Masiero, M. Pietroni and A. Riotto,
Nucl. Phys. {\bf B396}, 243 (1993);
M. Shirashi, Isao Umemura and K. Yamamoto,
Kyoto Preprint, NEAP-50 (1992).

\item{[11]}C.S. Aulakh and R.N. Mohapatra, Phys. Lett. {\bf B119},
136 (1983).

\item{[12]}
           J. Schechter and J.W.F. Valle,
        Phys. Rev. {\bf D25}, 774 (1982).

\item{[13]}E. Kolb and M. Turner, {\bf {The Early Universe}},
(Addison-Wesley 1990).

\item{[14]} T. Walker et al, Astrophys. J. {\bf 51}, 376 (1992).

\item{[15]}R. Scherrer and M. Turner, Astrophys. J. {\bf 331}, 19 (1988).

\bye